# Background-free and laser-ionization minimized single-shot fs-EFISH in luminous pulsed plasmas

**Vidyadhara B. V. Kumar,**[1] **Karna Patel,**[1,*] **Aman Satija,**[1] **Sally Bane,**[1,2] **and Terrence R. Meyer**[1,2,3]

[1] *School of Mechanical Engineering, Purdue University, West Lafayette, Indiana 47907, USA*
[2] *School of Aeronautics and Astronautics, Purdue University, West Lafayette, Indiana 47907, USA*
[3] *Department of Aerospace Engineering and Mechanics, University of Minnesota, Minneapolis, Minnesota 55455, USA*

**karna@purdue.edu*

**This work introduces spectrally tuned electric field induced second harmonic (t-EFISH) generation for substantially reducing or eliminating background plasma emission in single-shot, time-resolved, electric field measurements. The technique is demonstrated using a femtosecond 1240 nm pump and 620 nm signal detection to avoid spectral and temporal overlap with the early, intense onset of $N_2$(C→B) emission in nanosecond atmospheric-pressure air spark discharges. This enhances signal-to-emission ratios pre- and post-breakdown, enables lower pump energies, suppresses detector saturation, and reduces the coefficient of variation in the electric field measurements by 8.5x. This strategy also provides a pathway for background-free measurements in other discharge environments with varying spectro-temporal emission characteristics. Moreover, the longer 1240 nm pump wavelength reduces multiphoton intrusiveness, directly improving measurement fidelity.** 

Transient plasmas generated by nanosecond repetitively pulsed (NRP) discharges are used in a wide range of applications [1]. In these plasmas and other, the electric field is a fundamental property that governs plasma formation and is a key input parameter for validating plasma kinetics. Electric field induced second-harmonic (EFISH) generation is a non-resonant four-wave mixing technique that has gained substantial interest for the measurement of absolute electric field strength due to its sensitivity and temporal precision [2,3]. Steady-state and transient plasmas have been studied using EFISH with femtosecond (fs) [2], picosecond (ps) [3], and nanosecond (ns) [4] lasers. Since EFISH is a four-wave mixing process, these approaches can suffer from low signal levels. While ns and ps EFISH generally provide higher EFISH signal strengths, the advancement in fs EFISH is of interest due to its high repetition-rate, suitability for coupling with coherent Thomson and Raman diagnostics—enabling simultaneous, multiparametric plasma measurements—and the relatively high peak power at low pulse energy of fs lasers. This has the potential to reduce photolytic interference and unwanted scattering—which could plague ns and ps EFISH—but can lead to deviation from quadratic scaling at higher laser energies due to the occurrence of optical nonlinearities [5,6] such as self-phase modulation (SPM), Kerr self-focusing, and multiphoton ionization (MPI).

On-detector accumulation can increase the signal but reduce temporal precision in dynamic and stochastic environments. Hence, single-shot EFISH measurements using relatively lower laser energies are necessary for applications to transient plasmas, such as ns pulsed spark discharges. Due to the high luminosity of these discharges, however, the EFISH signal must be strong enough to overcome the intense background plasma emission on a single-shot basis. This includes emission from electronic transitions such as $N_2$(C→B), $N_2$(B→A), excited atomic species such as $O^+$, $N^+$, $NO^*$, and continuum [6–8]. Additionally, unlike ps and ns EFISH, where narrowband signals permit dispersive filtering of emission [3], fs EFISH has ~150 $cm^{-1}$ bandwidth, causing at least ~1000x poorer spectral isolation.

By shifting the EFISH pump wavelength using the tunable output from a fs optical parametric amplifier (OPA), we suppress background emission signal in the critical periods during the breakdown. It is found that these emission signals have different spectro-temporal evolution dynamics [6–8] enabling optimization of the signal-to-background ratio and precision of single-shot EFISH. Referred to here as tuned EFISH (t-EFISH, 1240 nm), this approach shifts the second-harmonic signal to ~620 nm, where the background emission was found to be minimal and delayed by ~8 ns relative to the conventional (c-EFISH, 800 nm) frequency-doubled 400 nm signal from a Ti:Sapphire laser. Hence, the electric field measurements using t-EFISH significantly reduce background and detector saturation that can overwhelm the EFISH signal. The reduction in single-shot noise and uncertainty enables accurate and repeatable single-shot measurements for understanding the breakdown dynamics and supporting model validation efforts.

The experimental setup is depicted in Fig. 1(a). A regeneratively amplified mode-locked Ti:Sapphire oscillator (fs-L, Solstice Ace) is the laser source, which produces 800 nm, ~100 fs pulses at 1 kHz. The system outputs ~7 mJ/pulse which is split into 3 mJ/pulse and 4 mJ/pulse beams using a beam splitter (BS). The 4 mJ/pulse portion pumps an OPA (Topas Prime) to generate a frequency-tuned beam at 1240 nm. The energy of the conventional 800 nm (c-EFISH) or tuned 1240 nm (t-EFISH) beams is controlled using a half-wave plate (HWP-1) and thin-film polarizer (TFP) combination. The beams are then passed through a long-pass filter (LPF-1) to remove any inherent second-harmonic present. A second half-wave plate (HWP-2) is used to optimize the polarization of the beam to be parallel to the externally applied electric field. Plano-convex lenses (L-1) with 300 mm focal length are used to focus and collimate the beam at the center between the electrodes. A dichroic mirror (DM) is used to separate and reflect the EFISH signal, and the beam is then passed through a series of filters (SPF, LPF-2, BPF) to remove any remaining stray signal. The EFISH signal is then focused on an ICCD camera (PIMAX4) using a 100 mm plano-convex lens (L-2). The beam transmitted through DM is directed to a photodiode (PD) after passing it through optical density (OD) filters. The details of the filters used are provided in the supplementary document. The ns pulsed plasma is generated in a 2 mm gap between two 0.5-inch diameter rounded tip stainless-steel electrodes (E). A DC power source is used for calibration, and a ns pulse generator (ns-PG, Eagle Harbor) for pulsed plasma EFISH experiments. The photodiode (PD), time-delay generator (TDG), and oscilloscope (OSC) are used to synchronize the laser with an ICCD and ns-PG. Voltage probes (P) and current probe (CP) are used to measure the voltage and current waveforms. All experiments are performed at atmospheric pressure and in quiescent air.

The similar beam size (~5 mm), energy (~80 μJ/pulse), electrode geometry, interelectrode gap (~2 mm), ICCD gate-width (3 ns), and gain (40) are maintained for all of the tests. An electrostatic simulation (Fig. 1(b)) is performed to obtain the path-averaged interelectrode Laplacian electric field [4]. Single-shot DC calibration (Fig. 1 (c, e)) is obtained for both of the EFISH configurations. The EFISH signal shows an expected quadratic dependence on applied electric field, with a sensitivity of ~2 kV/cm for both EFISH methods. This can be improved via on-detector accumulation but is not preferred for transient discharges to maintain fidelity. The DC calibration is followed by a time-varying voltage trace from the ns-PG. To ensure no breakdown, the peak voltage (~5 kV) is kept well below the breakdown voltage ($V_B$). As shown in Fig. 1(d, f), the EFISH data show good match with the voltage trace, as expected since the field remains Laplacian in the absence of breakdown. This match also validates the data collection and post-processing routine used for temporally resolved, single-shot fs EFISH.

To quantify electric field during breakdown using c-EFISH, the pulser is set to supply a peak voltage of ~25 kV with ~60 ns pulse duration. During data collection, an ensemble signal consisting of EFISH and plasma emission is collected, followed by plasma emission with the laser beam blocked. The EFISH signal is obtained by subtracting the plasma emission from the ensemble signal. The single-shot data points are averaged within 0.1 ns bins and converted to the corresponding electric field value. Fig. 2(a) shows the time-resolved electric field profile along with voltage and current traces. The time at which the electric field is maximum is defined as the point of breakdown. The ensemble signal ($I_{tot}$) and plasma emission ($I_e$) components are plotted in Fig. 2(b). The data in Fig. 2 (b) and Fig. 2 (e) shows that plasma emission begin to appear at ~18 ns, close to the breakdown, and shows substantial fluctuations thereafter. The signal-to-emission ratio (SER), in Fig. 2(c), represents $I_{tot}/I_e = (1 + I_{EFISH}/I_e)$ and shows that the background emission reaches the same magnitude as the EFISH signal (SER ~2 marked by the dash-dot line) near plasma breakdown at ~20 ns. At times, the SER drops below 2 when the total signal is dominated by plasma emission. This illustrates the challenge of EFISH measurements in highly luminous plasmas, especially in the spectral region contaminated by $N_2$(C→B) emission at ~400 nm, which appears at the onset [6–8] and obscures measurements.

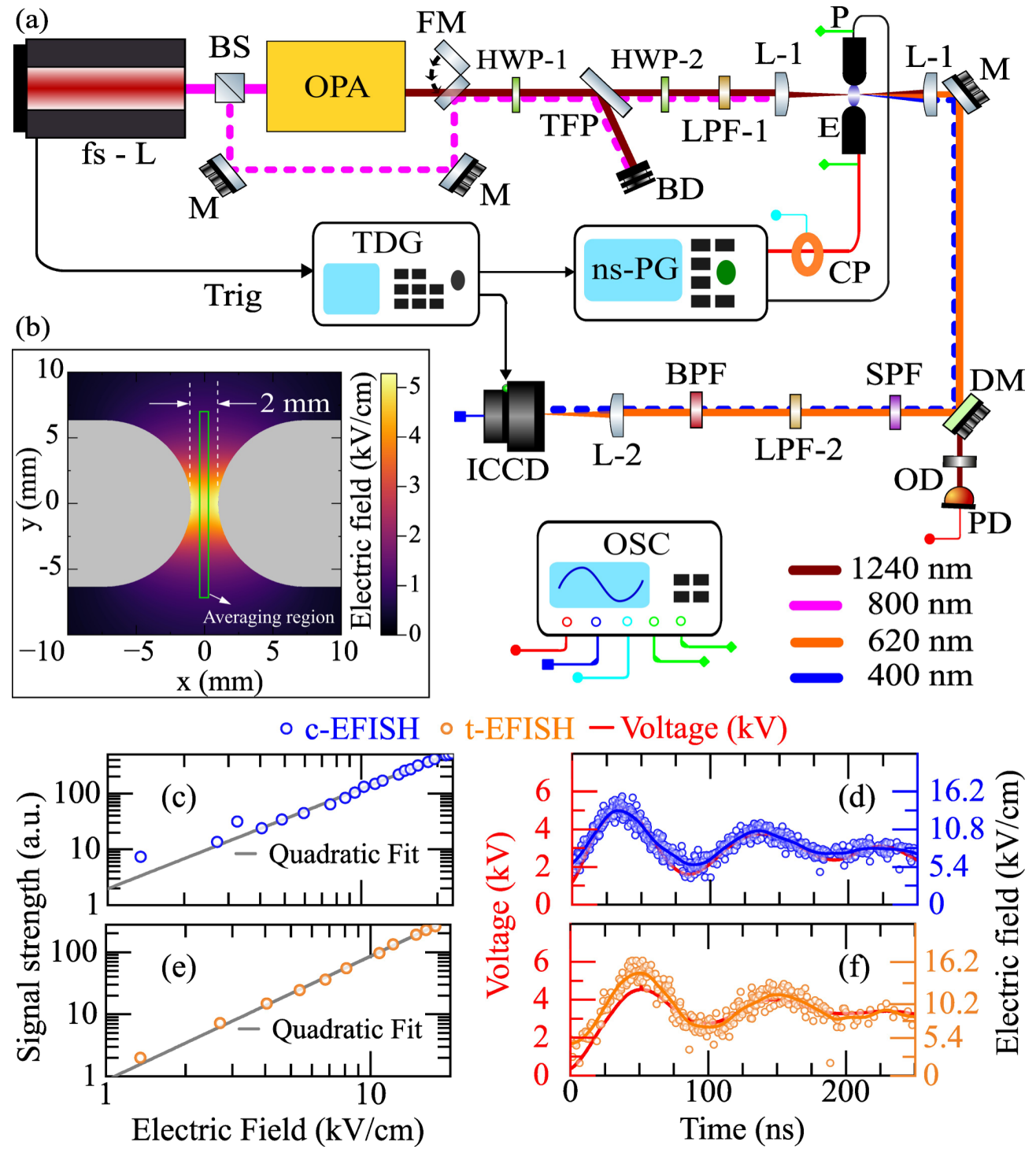

**Fig. 1.** (a) EFISH setup for 800 nm and 1240 nm pump beams. Dotted line represents conventional EFISH (c-EFISH, 800 nm) beam and solid line represents tuned EFISH (t-EFISH, 1240 nm) beam. (b) Electrostatic simulation for a 2 mm electrode gap at 1 kV applied potential. (c, e) Single-shot DC calibration data; each point is an average of 50 single-shots. (d, f) Single-shot time-resolved below-breakdown tests with a set voltage of 5 kV.

A statistical metric based on the coefficient of variation (COV) is used to quantify shot-to-shot variability (Fig. 2(d)). COV is defined as the ratio of standard deviation to the mean of the data, calculated separately for EFISH and emission. The ensemble signal and emission data points are allocated into 0.5 ns bins, over which the standard deviation and mean are calculated to determine the COV. The COV for c-EFISH before ~18 ns is less than 10% (Fig. 2, blue region). Near and beyond breakdown, the increasing and fluctuating emission COV propagates into the c-EFISH COV, compromising the reliability of the measurements.

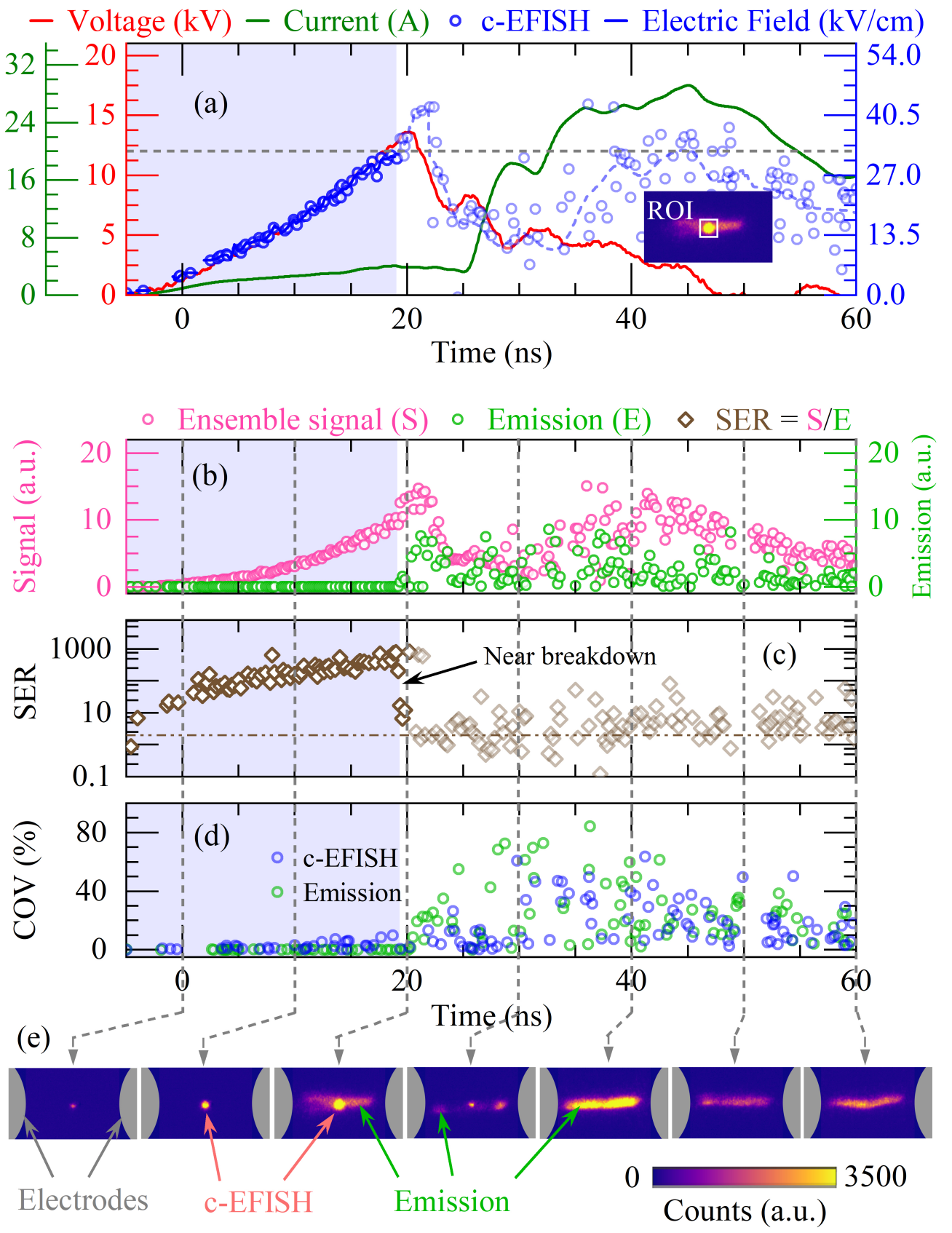

**Fig. 2**. Results of c-EFISH experiments. Time-resolved (a) electric field and (b) ensemble signal and plasma emission for breakdown test (set voltage ~25 kV). (c) Ensemble SER indicates influence of plasma emission on c-EFISH signal. (d) COV of c-EFISH and plasma emission. (e) ICCD images showing c-EFISH, background emission, and electrode location. Highlighted section (in blue) shows region where SER >2 and COV<10%.

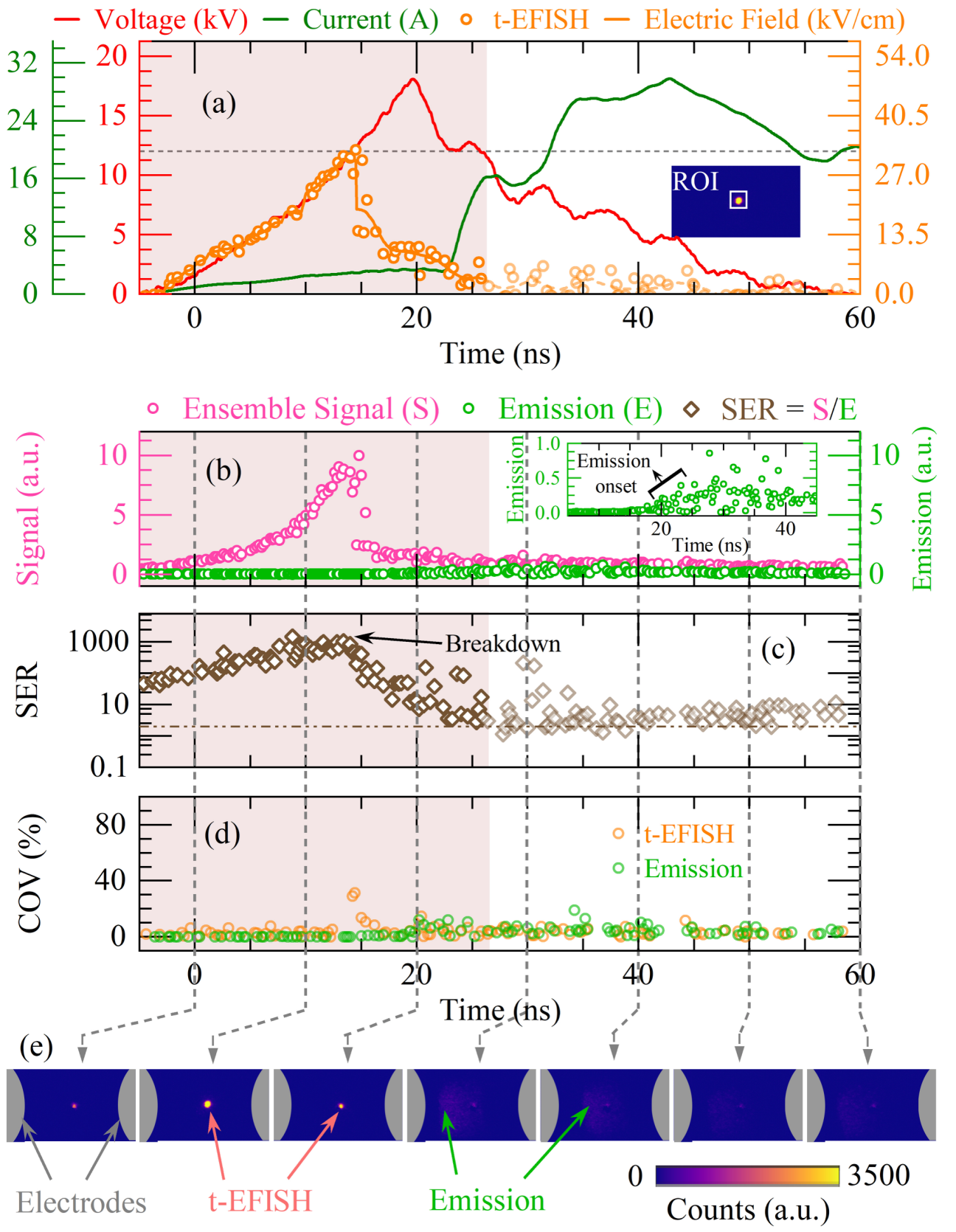

**Fig. 3.** Results of t-EFISH experiments. Time-resolved (a) electric field and (b) ensemble signal and plasma emission for breakdown test (set voltage ~25 kV). Inset image shows zoomed-in view of the emission onset. (c) Ensemble SER. (d) COV of t-EFISH and plasma emission. (e) ICCD images showing t-EFISH, background emission, and electrode location. Highlighted section (in red) shows the region where SER >2 and COV<10%.

To enable a rigorous test of plasma models near breakdown, it is important to separate the EFISH signal from the bright emission. This is accomplished by tuning the pump beam to 1240 nm to produce a t-EFISH signal at 620 nm—a spectral region between the dominant $N_2$(C→B) and $N_2$(B→A) emission manifolds—with low emission intensity that also beneficially exhibits a delayed onset [6–8]. The t-EFISH breakdown tests (Fig. 3) used the same voltage (~25 kV) and data collection procedure as c-EFISH. The measured field shows good agreement with the voltage trace until ~14 ns, after which it decreases rapidly (Fig. 3(a))—this deviation from the voltage trace was not observed for c-EFISH. However, this trend is consistent with prior work [4,9,10], which reported plasma breakdown experiments under different conditions in which the measured electric field deviates from the applied voltage and decreases before the voltage pulse peak. The steep decline in electric field at ~14 ns (Fig. 3(a)) denotes the onset of breakdown and implies that sufficient ionization has occurred. Voltage and displacement current increases until ~20 ns, with conduction current increasing at ~22 ns. We hypothesize that this ~8 ns delay is related to stabilization of the plasma sheath and finite-rate kinetics associated with secondary emission. Additional measurements of electron density and temperature would help validate this hypothesis. Notably, the breakdown (peak) electric field in Fig. 3(a) is ~31 kV/cm, similar to the peak field attained by c-EFISH prior to the emission onset (Fig. 2(a)).

Fig. 3(b) shows the time-resolved ensemble signal and plasma emission for t-EFISH. Unlike Fig. 2(b), the t-EFISH signal rises and drops without any background emission, since emission onset (inset, Fig. 3(b)) is delayed [6–8] and occurs after the EFISH measurement is acquired, and background emission intensity appears negligible even after onset. This validates the 620 nm region as a spectro-temporally 'quiet' region for EFISH measurements in air plasmas. Unlike c-EFISH (Fig. 2(c)), the t-EFISH SER stays above 2 for ~8 ns after breakdown, extending the post-breakdown measurement window (Fig. 3(c)). The COV in Fig. 3(d) remains near or below ~10% for the entire plasma pulse. A COV peak at breakdown (~14 ns) is expected due to the sudden EFISH signal drop within the 0.5 ns bin. Together, SER and COV indicate reliable t-EFISH measurements throughout

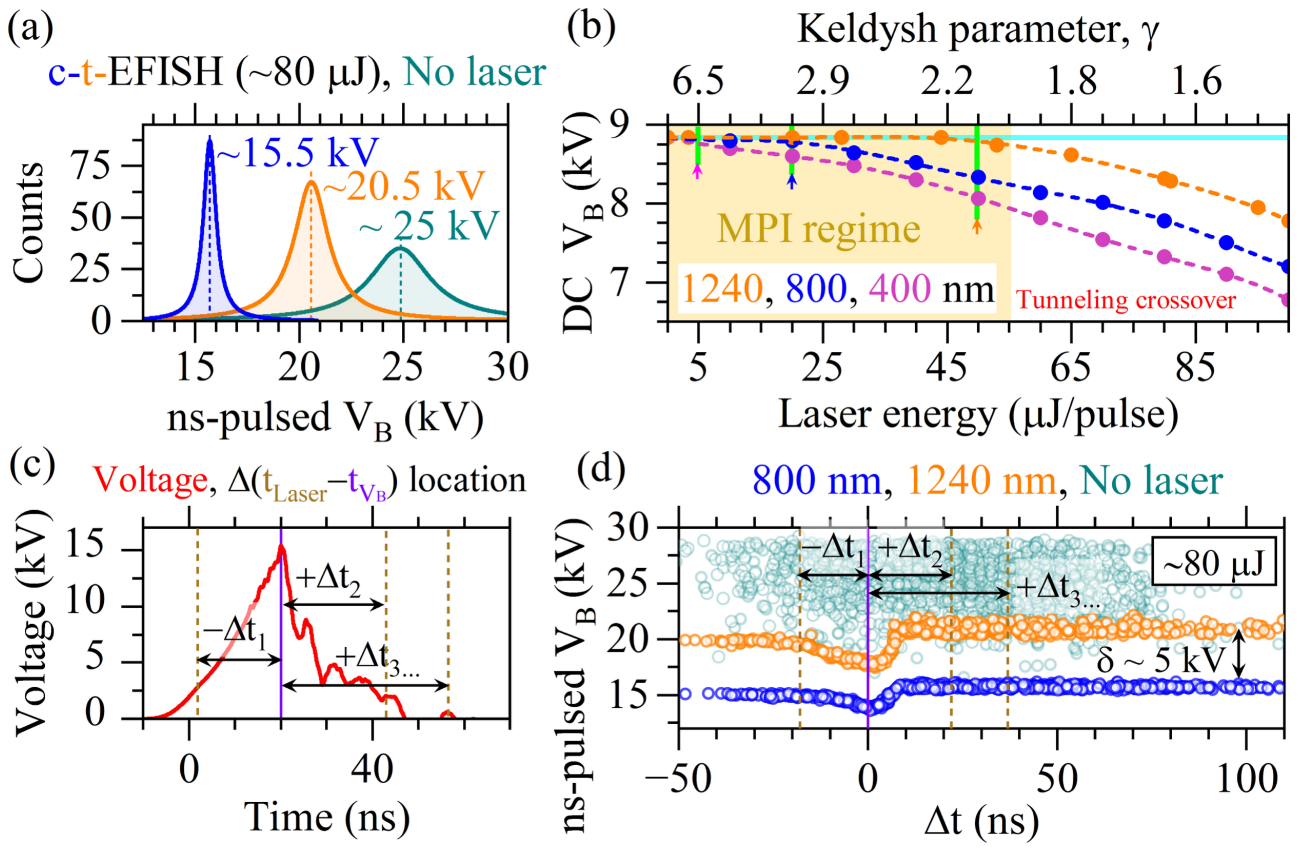

**Fig. 4**. (a) $V_B$ variation for no laser, c-EFISH, and t-EFISH, along with (b) systematic energy scaling across wavelengths. (c)-(d) Influence of the laser memory-effect on $V_B$. $\Delta t=\Delta(t_{Laser}-t_{VB})$ is the time-difference between laser pulse and plasma breakdown.

the field evolution, as shown by the red shaded region in Fig. 3.

Even at low 80 μJ/pulse energy, the high peak intensity of fs pulses can drive MPI, producing weak-ionization that promotes breakdown and lowers the $V_B$. To assess this, $V_B$ was recorded for no laser, t-EFISH (1240 nm), and c-EFISH (800 nm) (Fig. 4(a)). As expected, the unperturbed $V_B$ (~25 kV) dropped to ~20.5 kV for t-EFISH and ~15.5 kV for c-EFISH. This is likely due to lower photon energy at 1240 nm, which raises the MPI order (k) needed to ionize $O_2$, the dominant ionization source in air (~12.07 eV), from k=8 at 800 nm (~1.55 eV) to k=13 at 1240 nm (~1 eV). Because the MPI cross-section ($\sigma_k$) plummets by 10s of orders of magnitude with each increment in k, ionization is far less aggressive at 1240 nm, whereas 800 nm plausibly forms more seed electrons and lowers $V_B$ further. The tighter $V_B$ distribution at 800 nm also corroborates this, indicating laser-seeded electrons stabilized highly repeatable breakdown.

To fundamentally investigate laser-induced intrusiveness, we decoupled the electric pulse's temporal nature (over-voltage) by instead using a DC source (no-laser DC $V_B$ ~8.8 kV) and tracked $V_B$ against laser pulse energy across wavelengths (Fig. 4(b)). At 1240 nm (k=13), DC $V_B$ remains unchanged up to ~50 μJ/pulse, whereas shorter wavelengths drop sharply, branching off (color-coded arrows) at progressively lower energies of ~20 μJ/pulse for 800 nm (k=8) and ~5 μJ/pulse for 400 nm (k=4). Once the wavelength-dependent focal spot size is accounted for, these branch-off points fall at similar peak intensity (~9.5 TW/cm$^2$), which at first glance suggests a simple intensity-threshold process. However, this interpretation does not hold since the number of MPI electrons produced at 1240 nm and 800 nm are significantly lower compared to 400 nm at similar intensities due to the drastic drop in $\sigma_k$. We therefore hypothesize that geometric confinement and electron density together dictate the branching off onset. At 400 nm, a dense electron population is confined to a tight focal volume, whereas at 800 nm and 1240 nm even though likely far fewer electrons are generated, they are spread over a bigger space set by the longer Rayleigh range. Furthermore, DC $V_B$ for 1240 nm branches off at a similar intensity to 400 nm despite a much higher MPI order (k=13 vs. k=4), implying MPI likely isn't the major source for 1240 nm branch off since k=13 process would yield negligible MPI electrons at the k=3 onset intensity. This suggest that 1240 nm branch off is additionally influenced by a crossover to tunneling ionization, as indicated by the Keldysh parameter (γ) in Fig. 4 (b). Although theory places crossover at γ=1, a slight deviation is expected in experimental practices and was measured at γ~2 with our laser in prior work [11].

Additionally, kHz laser's 1 ms period is short, even at high atmospheric pressure, to prevent laser-induced memory effect [11]. This laser-generated (as ns pulsed plasma is operated once every 3 s to ensure no memory effect from it) buildup of residual charged species screen the background medium, lowering $V_B$, as inferred by the relative $V_B$ shift in Fig. 4(d). During laser-voltage pulse overlap (Fig. 4(c)), extra active species from ns-pulsed plasma breakdown lowers $V_B$ further, causing the sharp dip in Fig. 4(d). One subtlety is that at -Δt instances, the laser closely precedes the discharge, momentarily raising local seed-electron density (LSED) above the constant screened background and slightly lowering the $V_B$ baseline. At +Δt instances, however, the discharge has already occurred, so immediate laser-added LSED no longer lowers the baseline, leaving it slightly higher. Finally, memory effect is not an inherent drawback of high-speed fs lasers. It stems from residual MPI ionization, which can be mitigated by reducing pulse energy to the regime where the $V_B$ is unchanged (Fig. 4(b), 1240 nm), a focus for future work.

In summary, we demonstrate single-shot, time-resolved fs EFISH measurements in atmospheric pressure ns spark discharges. Spectral tuning from the conventional 800 nm to 1240 nm avoids the strong plasma emission that plagues measurements, yielding high SER, ~8.5x reduction in COV, and a strategy extendable to other gases and discharges. Influence of MPI and memory-effect is also studied, showing significantly reduced intrusiveness at a longer 1240 nm wavelength.

**Funding:** This work was supported by the U.S. Department of Energy, Office of Science, Office of Fusion Energy Sciences (Award No. DE-SC0021153).

**Disclosures:** The authors declare no conflicts of interest.

**Data availability.** Data underlying the results presented in this paper may be obtained from the authors upon reasonable request.

**Supplemental document.** See Supplement 1 for supporting content.

## REFERENCES

1. Y. Ju and W. Sun, Prog. Energy Combust. Sci. **48**, 21 (2015).
2. A. Dogariu, B. M. Goldberg, S. O'Byrne, et.al., Phys. Rev. Appl. **7**, 024024 (2017).
3. M. Z. Rahman, C. J. Kliewer, B. D. Patterson, et. al., Plasma Sources Sci. Technol. **35**, 015004 (2026)
4. I. V. Adamovich, T. Butterworth, T. Orriere, et.al., J. Phys. Appl. Phys. **53**, 145201 (2020).
5. T. Fujii, M. Sato, S. Nakamura, et.al., Opt. Lett. **46**, 238 (2021).
6. K. S. Patel, PhD thesis, Purdue University Graduate School (2024).
7. K. Patel, A. Saha, T. Zhou, et.al., Appl. Phys. Lett. **120**, 014101 (2022).
8. K. Patel, A. Saha, A. Satija, et.al., Appl. Phys. Lett. **129**, 064101 (2026).
9. N. D. Lepikhin, D. Luggenhölscher, U. Czarnetzki, J. Phys. Appl. Phys. **54**, 055201 (2020).
10. T. Ito, T. Kanazawa, S. Hamaguchi, Phys. Rev. Lett. **107**, 065002 (2011).
11. A. Sharma, M. N. Slipchenko, M. N. Shneider, et. al., Sci. Rep. **8**, 2874 N. (2018).

## FULL REFERENCES

1. Ju, Y., & Sun, W. (2015). Plasma assisted combustion: Dynamics and chemistry. *Progress in Energy and Combustion Science*, *48*, 21–83. https://doi.org/10.1016/j.pecs.2014.12.002

2. Dogariu, A., Goldberg, B. M., O'Byrne, S., & Miles, R. B. (2017). Species-Independent Femtosecond Localized Electric Field Measurement. *Physical Review Applied*, *7*(2), 024024. https://doi.org/10.1103/PhysRevApplied.7.024024

3. Rahman, M. Z., Kliewer, C. J., Patterson, B. D., & Jiang, C. (2026). Determining the electric field in a 10 ns pulsed plasma in fuel-air mixtures using EFISH. Plasma Sources Science and Technology, 35(1), 015004. https://doi.org/10.1088/1361-6595/ae2efd

4. Adamovich, I. V., Butterworth, T., Orriere, T., Pai, D. Z., Lacoste, D. A., & Cha, M. S. (2020). Nanosecond second harmonic generation for electric field measurements with temporal resolution shorter than laser pulse duration. *Journal of Physics D: Applied Physics*, *53*(14), 145201. https://doi.org/10.1088/1361-6463/ab6790

5. Fujii, T., Sato, M., Nakamura, S., Kumada, A., Miki, M., & Oishi, Y. (2021). Electric-field-induced second-harmonic generation using high-intensity femtosecond laser pulses over the observable optical breakdown threshold. *Optics Letters*, *46*(2), 238. https://doi.org/10.1364/OL.412856

6. Patel, K. S. (2024). *Ultrafast Emission Spectroscopy and Nonlinear Laser Diagnostics for Nanosecond Pulsed Plasmas* [Thesis, Purdue University Graduate School]. https://doi.org/10.25394/PGS.25670793.v1

7. Patel, K., Saha, A., Zhou, T., Meyer, T. R., Bane, S., & Satija, A. (2022). Spectrally filtered ps–ns emission dynamics of atmospheric-pressure nanosecond pulsed plasmas. *Applied Physics Letters*, *120*(1), 014101. https://doi.org/10.1063/5.0073630

8. Patel, K., Saha, A., Satija, A., Meyer, T. R., & Bane, S. (2026). Single-shot ultrafast dynamics of nanosecond pulsed plasmas: Transition from ps–ns nonequilibrium to near-full ionization. *Applied Physics Letters*, *129*(6). https://doi.org/10.1063/5.0341587

9. Lepikhin, N. D., Luggenhölscher, D., & Czarnetzki, U. (2020). Electric field measurements in a He:N2 nanosecond pulsed discharge with sub-ns time resolution. *Journal of Physics D: Applied Physics*, *54*(5), 055201. https://doi.org/10.1088/1361-6463/abbbb4

10. Ito, T., Kanazawa, T., & Hamaguchi, S. (2011). Rapid Breakdown Mechanisms of Open Air Nanosecond Dielectric Barrier Discharges. *Physical Review Letters*, *107*(6), 065002. https://doi.org/10.1103/PhysRevLett.107.065002

11. Sharma, A., Slipchenko, M. N., Shneider, M. N., Wang, X., Rahman, K. A., & Shashurin, A. (2018). Counting the electrons in a multiphoton ionization by elastic scattering of microwaves. *Scientific reports*, *8*(1), 2874. https://doi.org/10.1038/s41598-018-21234-y